\documentclass[preprint,prd,showpacs,superscriptaddress,preprintnumbers,nofootinbib,fleqn,floatfix]{revtex4}

\usepackage{amsmath}
\usepackage{amssymb}
\usepackage{bm}
\usepackage{dcolumn}
\usepackage{graphicx}
\usepackage{rotating}
\usepackage{subfigure}
\usepackage{axodraw}
\usepackage{longtable}
\usepackage{color}
\usepackage{enumerate}

\newcommand{\del}{\partial}

\newcommand{\iu}{{\rm i}}

\newcommand{\beq}{\begin{equation}}
\newcommand{\eeq}{\end{equation}}
\newcommand{\bea}{\begin{eqnarray}}
\newcommand{\eea}{\end{eqnarray}}
\newcommand{\bsub}{\begin{subequations}}
\newcommand{\esub}{\end{subequations} \noindent}

\def\drv#1{{\partial_{#1}}} 
\def\drvstar#1{\partial\kern-0.5pt\smash{\raise 4.5pt\hbox{$\ast$}}
               \kern-5.0pt_{#1}} 
\def\lvec#1{\setbox0=\hbox{$#1$}
    \setbox1=\hbox{$\scriptstyle\leftarrow$}
    #1\kern-\wd0\smash{
    \raise\ht0\hbox{$\raise1pt\hbox{$\scriptstyle\leftarrow$}$}}
    \kern-\wd1\kern\wd0} 
 
\def\ldrvstar#1{\lvec{\,\partial}\kern-0.5pt\smash{\raise 4.5pt\hbox{$\ast$}}
               \kern-5.0pt_{#1}} 

\begin{document}

\date{\today}


\title{%
 Positronium resonance contribution to the electron $g-2$
}

\author{Masashi Hayakawa}
\affiliation{Department of Physics, Nagoya University, Nagoya, Japan 464-8602 }
\affiliation{Theoretical Physics Laboratory, Nishina Center, RIKEN, Wako, Japan 351-0198 }

\begin{abstract}
 Recently a few authors pointed out that 
the positroniums give rise to
an extra contribution to the electron $g-2$
which is independent of the  
perturbative calculation up to $O(\alpha^5)$ 
and has the same magnitude as the $O(\alpha^5)$ perturbative effect. 
 Here, we scrutinize how the positronium 
resonances contribute to the electron $g-2$
through the vacuum polarization function, 
and conclude that there is no additional sizable $O(\alpha^5)$ 
contribution from the positronium resonances to the electron $g-2$.
\end{abstract}


\maketitle

\section{Introduction}

 Recently, Ref.~\cite{Mishima:2013ama} pointed out
that 
the positroniums in the vector channel
give an additional contribution to the electron $g-2$, $a_e$,
which cannot be captured by the perturbative analysis 
up to $O(\alpha^5)$ \cite{Aoyama:2012wj}. 
 Subsequently, Ref.~\cite{Fael:2014nha} checked 
the calculation in Ref.~\cite{Mishima:2013ama} 
and presented an updated value for such a contribution.
 They concluded that 
the correction is independent of the perturbative contribution 
and has the same order of magnitude
as the $O(\alpha^5)$ perturbative contribution
and thus affects to the comparison of 
the experiment and the theory of $a_e$.

 The assertion in Refs.~\cite{Mishima:2013ama,Fael:2014nha} 
seems to be gradually attaining the consensus
in the community of particle phenomenology.
 During the preparation of this article,
the two papers \cite{Melnikov:2014lwa,Eides:2014swa} 
presented a negative conclusion on 
the results in Refs.~\cite{Mishima:2013ama,Fael:2014nha}.
 In such a circumstance, this article attempts to scrutinize
the current issue from the basic of the quantum field theory.
 The consideration in full order QED in Sec.~\ref{sec:positro} shows that
Refs.~\cite{Mishima:2013ama,Fael:2014nha}
do not dealt with the contribution of the positronium resonances.
 The proper identification of such a contribution
immediately shows that 
there is no contribution to $a_e$ from the positronium resonances
with the size found in Refs.~\cite{Mishima:2013ama,Fael:2014nha}.

 In Sec.~\ref{sec:positro}, 
we start with summarizing the question to be addressed here
and present the answer to it.
 Section~\ref{sec:discussion} discusses
the connection of this paper with those of the precedence works 
\cite{Mishima:2013ama,Fael:2014nha,Melnikov:2014lwa,Eides:2014swa}.
 It turns out that 
the analysis perspective itself, 
which provides a more convincing approach to the question,  
is quite different from
that in 
Refs.~\cite{Mishima:2013ama,Fael:2014nha,Melnikov:2014lwa,Eides:2014swa}.

\section{Positronium contribution}
\label{sec:positro}

\begin{figure}[htb]
\includegraphics[scale=1.0,clip]{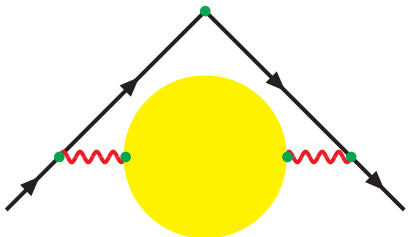}
\caption{Contribution to the electron $g-2$
from the vacuum polarization function induced by QED (the blob part).
 The black and wavy red lines denote
the propagation of electron and photon, respectively. 
}
\label{fig:g-2fromVP}
\end{figure}

 Following Ref.~\cite{Mishima:2013ama}, we restrict our attention
to the type of the QED contribution 
induced through the vacuum polarization to the electron $g-2$
as shown in Fig.~\ref{fig:g-2fromVP}.

 In order to disentangle the confusion, 
we first marshal the question itself to be addressed here\,: 
How large is the contribution from the positroniums in the vector channel 
to the electron $g-2$ through Fig.~\ref{fig:g-2fromVP}.

 To this end, we start with reconsidering 
{\it full-order} QED contribution to the two-point function 
of the electromagnetic current 
$j_\mu \equiv -e\,\overline{\psi} \gamma_\mu \psi$ 
in the QED with the electron only, which suffices 
for the succeeding discussion
\begin{align} 
 &
 \int d^4 x\,e^{\iu q \cdot x}\, 
 \left<0\right|T j_\mu(x) j^\nu(0)\left|0\right>_{\rm 1PI} 
 =
 \iu \left(\delta_\mu^{\ \nu} q^2 - q_\mu q^\nu\right) \Pi(q^2)\,.
  \label{eq:Def:VP}
\end{align}
 The renormalized function 
will be obtained by $\Pi_{\rm R}(q^2) = \Pi(q^2) - \Pi(0)$.

 Since QED does not have any nontrivial classical gauge configurations 
such as instantons, we can identify which set of Feynman diagrams, 
{\rm e.g.} an infinite series of ladder diagrams,
is associated with the quantum dynamics 
relevant to the phenomenon of one's interest.
 
 It is worthwhile to recall 
some basic features of the state space and $\Pi(q^2)$.
 The physical space of QED is spanned by stable one-particle states 
and multi-particle states composed of them.
 Since confinement does not occur and 
no stable bound state exists in QED,
the only one-particle states are photon, electron
and positron. 
 Every multi-particle state consists of photon(s) and electron(s).

 The vacuum polarization function $\Pi(q^2)$ defined in 
Eq.~(\ref{eq:Def:VP}) is analytic on the surface
obtained from two complex planes by braiding on the branch cuts.
 Each of the branch cuts is associated with a multi-particle state
$\left|\Psi;\,\left\{{\bf q}_j,\,\lambda_j\right\}_j\right>$
($\lambda_j$ denotes the polarization.)
that couples non-trivially to the electromagnetic current $j_\mu$; 
$\left<0\right| j_\mu(0)
 \left|\Psi;\,\left\{{\bf q}_j,\,\lambda_j\right\}_j\right> \ne 0$.
 The examples of such multi-particle states are 
multi-photons,  $3 \gamma$, $5 \gamma$, 
or $e^- e^+$, $e^- e^+ \gamma$, etc.
 The kinematics involved in 
the matrix element, say, 
$\left<0\right| j_\mu(0)
 \left|3\gamma;\,\left\{{\bf q}_j,\,\lambda_j\right\}_{j=1,2,3}\right>$
can be found in Ref.~\cite{Costantini:1971cj}.
 With this analytic structure of $\Pi(q^2)$ in our mind, 
we derive the dispersion relation for $\Pi_{\rm R}(q^2) /q^2$ 
after introducing the infrared regulator 
so that the branch cuts of multi-photons start
from infinitesimally small constant $s_0 > 0$ 
\begin{align}
 &
 \frac{\Pi_{\rm R}(q^2 + \iu \epsilon)}{q^2 + \iu\,\epsilon}
 =
 -
  \frac{1}{\pi}
  \int_{0+}^\infty \frac{ds}{s}\,
  \frac{{\rm Im}\,\Pi_{\rm R}(s + \iu\,0)}{q^2 - s + \iu \epsilon}
 \,. \label{eq:VPdispersionRelation}
\end{align}
 This together with Eq.~(\ref{eq:Def:VP}) 
immediately yields the expression for
the contribution to $a_e$ of the type in FIG.~\ref{fig:g-2fromVP}
as a superposition of 
the contribution $a_e(s)$ from the vector boson with mass squared $s$
weighted by ${\rm Im}\,\Pi_{\rm R}(s + \iu\,0)$ 
\begin{equation}
 a_e[{\rm vp}] = 
 \int_{0+}^\infty \frac{ds}{s}\,
 {\rm Im}\,\Pi_{\rm R}(s + \iu\,0)\,a_e(s)\,.
  \label{eq:ae_massiveVector}
\end{equation}

 In fact, the branch cuts associated with the multi-photons 
are overlooked in FIG.~3 of Ref.~\cite{Mishima:2013ama}.
 Instead, $\Pi(q^2)$ is supposed to have complex poles.
 However, complex poles are just the concepts
that are often {\it introduced temporarily
in particle phenomenology}
for the  purpose to calculate the total decay width and 
make comparison with the experiments.
 The imaginary part of a complex pole, 
the decay width, depends upon one's definition.
 The requirement of the gauge independence, for instance,
may motivate to choose a more favorable one 
\cite{Sirlin:1991rt}.
 {\it Theoretically}, the unitarity is assured 
only if $\Pi(q^2)$ can receive nontrivial contribution from 
the states $\left|\Psi;\,\left\{{\bf q}_j,\,\lambda_j\right\}_j\right>$
such that
$\left<0\right| j_\mu(0)
 \left|\Psi;\,\left\{{\bf q}_j,\,\lambda_j\right\}_j\right>
\ne 0$, 
which results in producing the branch cuts of $\Pi(q^2)$.

\begin{figure}[htb]
\includegraphics[scale=1.0,clip]{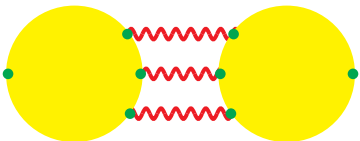}
\caption{A typical positronium contribution to
the vacuum polarization function.
 The blob part denotes 
the one-particle irreducible correlation function 
of four electromagnetic currents. 
}
\label{fig:positroniumEffectVP}
\end{figure}

 Now, Eq.~(\ref{eq:ae_massiveVector}) together with
the above remark on the analytic property of $\Pi_{\rm R}(q^2)$
immediately allows to identify which type of diagrams 
contains the positronium contribution.
 The ortho-positronium contribution 
is associated with ${\rm Im}\,\Pi_{\rm R}(s)$
originating from $(2n + 3) \gamma$ states ($n = 0,\,1,\,2,\cdots$).
 This is because 
the component involved in 
the state $j_\mu(0) \left|0\right>$ 
that can be considered as the positronium ground state, say, 
should be concentrated in the region centered at 
$\sqrt{s} \simeq 2 m_e - \alpha^2 m_e /4$ with the narrow width
$\propto \alpha^6 m_e$ 
so that its overlap with
each multi-state containing $e^- e^+$ is vanishing small.
 Therefore, the {\it physical} ortho-positronium resonance contribution
comes from, say, the diagram in FIG.~\ref{fig:g-2fromVP}, 
but with the vacuum polarization part 
replaced by a particular form shown in FIG.~\ref{fig:positroniumEffectVP}, 
which is distinctly different than that dealt with 
in Refs.~\cite{Mishima:2013ama,Fael:2014nha} 
and essentially that in Refs.~\cite{Melnikov:2014lwa,Eides:2014swa}.
 Such diagrams at the leading-order will be 
$O(\alpha^7)$ if at least one photon must be exchanged in each of 
the two light-by-light scattering amplitudes
in FIG.~\ref{fig:positroniumEffectVP} to form a positronium.
 The smallness of such a contribution 
will be speculated from 
the result \cite{Aoyama:2008gy} for the $O(\alpha^5)$ contribution
to $a_e$ caused by the diagrams of the type
in FIG.~\ref{fig:positroniumEffectVP} with 
the one-loop light-by-light scattering subdiagrams, which belong to Set I(j) 
according to the classification scheme of $O(\alpha^5)$ diagrams 
\cite{Aoyama:2005kf,Kinoshita:2005sm}
\begin{equation}
 a_e[{\rm I(j)}, e\ {\rm only}] 
 = 0.000\ 3950\ (87) \left(\frac{\alpha}{\pi}\right)^5\,.
  \label{eq:ae_Ij:e_only}
\end{equation}
 This is quite smaller than the dominant tenth-order contribution
which is found to have magnitude $O(1) \times (\alpha/\pi)^5$
\cite{Aoyama:2012wj}.
 The resonance contribution starting at $O(\alpha^7)$
will be further suppressed by the factor $\left(\alpha /\pi\right)^2$.

 In this paper, we see how the physical positronium resonance contributes 
to the electron $g-2$ through the diagram in
FIG.~\ref{eq:VPdispersionRelation}.
 Needless to say, there is no systematic way to single out 
the resonance contribution by separating it from the continuum contribution.
 Properly speaking,
what is done above is to correctly identify 
a set of the diagrams which contains the contribution of positronium
resonances in the vector channel. 

\section{Conclusion and discussion}
\label{sec:discussion}

 The plain consideration in Sec.~\ref{sec:positro} 
enabled us to correctly identify a set of diagrams
that contain the positronium resonance contribution, 
and leads the conclusion that there is no 
additional nonperturbative correction to the electron $g-2$
of the same size as the perturbative correction of $O(\alpha^5)$ 
as was first pointed out in Ref.~\cite{Mishima:2013ama}.
 Here, we discuss the connection of our founding 
with the precedence works concerning with the current issue.

 Obviously, the difference of this work
from Refs.~\cite{Mishima:2013ama,Fael:2014nha,Melnikov:2014lwa,Eides:2014swa}
stems from the fact that we never neglect the instability of 
the positroniums and deal with the proper space of states in QED.
 The question to be addressed in this article is defined
at the beginning of Sec.~\ref{sec:positro}, 
and we obtain a definite answer to it.
 In contrast, the other works seems to cast the following question 
by neglecting the unstable character of positroniums:
 Is the QED dynamics which are mainly concerned with
the formation of positroniums give rise 
to an additional nonperturbative contribution to the electron $g-2$ ?

 Instead of chasing the details of the discussions in 
Refs.~\cite{Mishima:2013ama,Fael:2014nha,Melnikov:2014lwa,Eides:2014swa}, 
we discuss the following points in the rest of the paper:
\begin{itemize}
%
\item
On one hand, one focuses on some coulombic dynamics nonperturbatively.
On the other hand, one wishes to forbid the decay 
of the bound state, 
which cannot be realized just from the perturbative order counting.
We examine theoretically
what concrete approximation reconciles
these seemingly contradictory situations.
\item
There is no local field theory
that reproduces only the approximation to the two-point function, 
i.e. the connected diagram contribution.
It is thus inevitable to deal with another types of contribution 
which unstabilize positroniums, 
and to examine the current issue by working with the state space
as described in Sec.~\ref{sec:positro}.
\end{itemize}


\begin{figure}[htb]
\includegraphics[scale=1.0,clip]{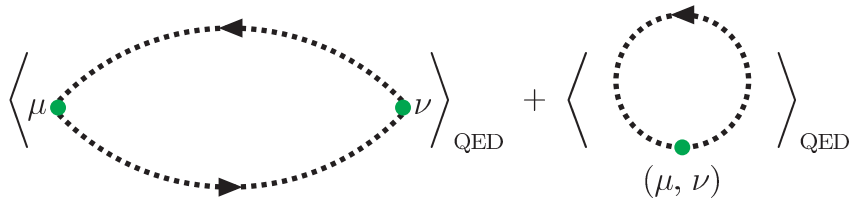}
\caption{
 Connected diagram contribution to
the two point function of the electromagnetic currents 
in lattice QED.
 An arrowed dotted line represents
the inverse of the electron Dirac operator,
$D[A]^{-1}(x,\,y)$.
 $\left<\mathcal{Q}\right>_{\rm QED}$ denotes
the vacuum expectation value of the quantity $\mathcal{Q}$ in QED.
 The second term contributes only if the positions 
of the two current operators coincide with each other.
}
\label{fig:ConnDiagLatticeQED}
\end{figure}

 We adopt the lattice regulariation for QED
just because the terminology in the framework of the lattice field theory 
is used in the following discussion.
 Figure \ref{fig:ConnDiagLatticeQED}
corresponds to 
the (gauge-invariant) nonperturbative approximation
to the vacuum polarization function (\ref{eq:Def:VP})
taken in
Refs.~\cite{Mishima:2013ama,Fael:2014nha,Melnikov:2014lwa,Eides:2014swa}.
 Each arrowed dashed line in that figure
is not the fermion propagator in the perturbation theory, 
but the inverse of the Dirac operator of the electron, 
$D[A]^{-1}(x,\,y)$, under a given gauge potential $A$.
 The symbol $\left<\mathcal{Q}[A]\right>_{\rm QED}$ denotes 
the vacuum expectation value (VEV) of a quantity $\mathcal{Q}[A]$
that depends only on the gauge potential in QED
\footnote{
It is not necessary to work in Euclidean space 
unless one attempts to simulate the system.
}
\begin{align}
 \left<\mathcal{Q}[A]\right>_{\rm QED} =&\,
 \frac{1}{Z_{\rm QED}}
 \int d\psi\,d\overline{\psi}\,dA\,
 \exp\left(\iu S_{\rm QED}\left[A,\,\psi,\,\overline{\psi}\right]\right)
 \mathcal{Q}[A] \nonumber\\
 =&\,
 \frac{1}{Z_{\rm QED}}
 \int dA\,
 \exp\left(
  \iu S_{\rm G}[A]
 \right)
 {\rm det}\left(- D[A]\right)\,
 \mathcal{Q}[A] \,,\nonumber\\
 S_{\rm QED}\left[A,\,\psi,\,\overline{\psi}\right] =& S_{\rm G}[A]
 + a^4 \sum_{x} \overline{\psi}(x) (\iu D[A] \psi)(x)\,,\nonumber\\
 S_{\rm G}[A]
 =&\,  
 a^4 \sum_{x} 
 \left\{
  - \frac{1}{4}\,F_{\mu\nu}(x)\,F^{\mu\nu}(x)
  + \frac{1}{2\lambda} \left(\drvstar{\mu} A^\mu\right)^2
 \right\} \,,
  \nonumber\\
 Z_{\rm QED} =&
 \int d\psi\,d\overline{\psi}\,dA\,
 \exp\left(\iu S_{\rm QED}\left[A,\,\psi,\,\overline{\psi}\right]\right)\,,
\end{align}
where``$a$'' denotes the lattice spacing, 
$F_{\mu\nu}(x)$ in the noncompact formulation of the lattice
QED takes the familiar form 
$F_{\mu\nu}(x) = \del_\mu A_\nu(x) - \del_\nu A_\mu(x)$
but with the forward difference 
$a\,\drv{\mu} f(x) \equiv f(x + a \widehat{\mu}) - f(x)$, 
and $\drvstar{\mu}$ denotes the backward difference operator; 
$a\,\drvstar{\mu} f(x) \equiv f(x) - f(x - a \widehat{\mu})$.
 The tadpole diagram in FIG.~\ref{fig:ConnDiagLatticeQED}
appears because $D[A]$ involves the 
Wilson line, $e^{-\iu e\,a A_\mu(x)}$, which parallel-transports back
the variable at $x + a \widehat{\mu}$ to $x$. 
 
\begin{figure}[htb]
\includegraphics[scale=1.0,clip]{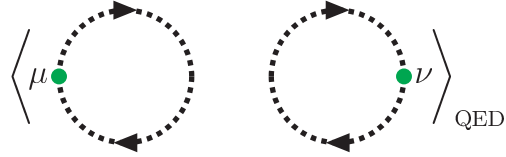}
\caption{
 Disconnected diagram contribution to
the correlation function of two electromagnetic currents. 
 The diagram in FIG.~\ref{fig:positroniumEffectVP} appears
as a part of this type of diagrams.
}
\label{fig:DscDiagLatticeQED}
\end{figure}


 Noting that the VEV of the operator 
$O\left[\psi,\,\overline{\psi},\,A\right]$ depending
on the electron fields can be converted to that of 
$\widehat{O}[A]$ defined by 
\begin{equation}
 \widehat{O}[A] \equiv
 \left[
  O\left[
   \frac{1}{a^4} \frac{\del^L}{\iu \del \overline{\eta}},\,
   \frac{1}{a^4} \frac{\del^R}{\iu \del \eta},\,A
  \right]
  \exp\left(
   a^4 \sum_x a^4 \sum_y
   \iu \overline{\eta}(x)\,D[A]^{-1}(x,\,y)\,\iu \eta(y)
  \right)
 \right]_{\eta,\,\overline{\eta} \rightarrow 0}\,,
\end{equation}
with the left (right) derivative $\del^L /\del \overline{\eta}(x)$ 
($\del^R /\del \eta(y)$), 
the total contribution to 
the full correlation function of two electromagnetic currents
\begin{align} 
 &
 \int d^4 x\,e^{\iu q \cdot x}\, 
 \left<0\right|T j_\mu(x) j^\nu(0)\left|0\right>
 =
 \iu \left(\delta_\mu^{\ \nu} q^2 - q_\mu q^\nu\right) J(q^2)\,,
  \label{eq:Def:VP_full}
\end{align}
is found to be given by 
the sum of the connected diagram in FIG.~\ref{fig:ConnDiagLatticeQED} and 
the {\it disconnected} diagram shown in FIG.~\ref{fig:DscDiagLatticeQED}, 
which also contains the contribution of the one-particle reducible diagrams.
 A simple diagrammatic consideration enables to express $J(q^2)$ 
in term of $\Pi(q^2)$
\begin{equation}
 J(q^2) = \frac{\Pi(q^2)}{1 + \Pi(q^2)}\,.
 \label{eq:full_1PI}
\end{equation}
 We recall that the disconnected diagram
contains the diagrams responsible to the decay of the positroniums.

 The simulation of the connected diagram 
in FIG.~\ref{fig:ConnDiagLatticeQED} will allow to measure 
the masses of the {\it pseudo}-bound states in the vector channel 
that are absolutely stable.
 However, the connected diagram contains, say, the ladder-type photon exchange 
only in the ``$t$-channel'', 
where the``$s$-channel'' is taken to be in the direction of the injected 
momentum in FIG.~\ref{fig:ConnDiagLatticeQED}.
 If we cut the subdiagrams with four external fermion lines 
out of a perturbative diagram of the type in FIG.~\ref{fig:ConnDiagLatticeQED} 
and embed it again into the rest of the original
after rotating it by $90$ degrees in a clockwise
direction, 
we will obtain a diagram of the type in FIG.~\ref{fig:DscDiagLatticeQED}.
 This indicates that 
there is no local Lagrangian density 
that reproduces the contribution from the connected diagram 
and no contribution from disconnected diagram.
 
 The situation should be contrasted with the case 
in which 
the decay process caused by the weak interaction is neglected.
 Then, there exists a local field theory that describes
the system with the weak interaction switched off
\footnote{
 The weak interaction will be switched off 
by letting the VEV of Higgs doublet $v$ 
and the electron yukawa coupling $y_e$ going to $0$ 
with the electron mass $m_e = y_e v /\sqrt{2}$ fixed finite.
},
and we can construct the state space at the zeroth-order of 
the approximation.
 A more concrete and familiar example is the description of hadron physics
where the zeroth-order is approximated by the world with QCD only 
and the corrections due to QED 
\footnote{
The electromagnetic correction to the meson masses
can also be incorporated perturbatively 
as in Ref.~\cite{deDivitiis:2013xla}.
}
and the dynamics of weak interactions are managed perturbatively.
 If no local field theory describing the zeroth-order approximation exists, 
we cannot proceed with the calculation of $a_e$ 
with use of the dispersion relation as Eq.~(\ref{eq:ae_massiveVector}) 
which relies on the analytic property of $\Pi(q^2)$, 
the existence of the state space and unitarity. 
 Hence, we have to tackle with the current issue using 
the state space described as in Sec.~\ref{sec:positro}
and Eq.~(\ref{eq:ae_massiveVector}).
 The connected diagram 
in FIG.~\ref{fig:ConnDiagLatticeQED} gives a significant contribution
to the electron $g-2$.
 Eq.~(\ref{eq:ae_massiveVector}) implies that 
such a contribution comes from 
the intermediate states of $e^- e^+$,
$e^- e^+ \gamma$, etc.~and can be calculated by means of perturbation.

\begin{acknowledgements}
 The author thanks G.~Mishima for 
letting him know full details of Ref.~\cite{Mishima:2013ama}
and the basic of Bethe-Salpeter amplitude.
\end{acknowledgements}


\end{document}